\documentclass[12pt]{article}
\def\be{\begin{equation}}
\def\ee{\end{equation}}
\def\ba{\begin{eqnarray}}
\def\ea{\end{eqnarray}}

\def\nn{\nonumber}
\begin{document}
\title{Extensivity and nonextensivity of two-parameter entropies}
\author{S. Asgarani \thanks{email: sasgarani@ph.iut.ac.ir} ,
       \ \ B. Mirza \thanks{email: b.mirza@cc.iut.ac.ir} \\ \\
{\it Department of  Physics, Isfahan University of Technology (IUT)}\\
{\it Isfahan,  Iran,} \\}

\date{}
\maketitle{$\hspace{5cm}$\Large{Abstract}}\\
In this paper, we investigate two-parameter entropies and obtain
some conditions for their extensivity. By using a generalized
$(k,r)-product$, correlations for subsystems are related to the
joint probabilities, so that the entropy remains extensive.
\newpage
\section{Introduction}
A quantity $X(A)$ associated with a system $A$ is said additive
with
regard to a specific composition of $A$ and $B$ if it satisfies\\
\be\label{01} X(A+B)=X(A)+X(B) \ee  where + inside the argument
of $X$ precisely indicates that composition. suppose, instead of
two subsystems $A$ and $B$, we have $N$ of them
$(A_{1},A_{2},...,A_{N})$. Then the quantity $X$ is additive if we
have \be\label{02} X({\sum_{i=1}^n}A_{i})={\sum_{i=1}^n}X(A_{i})
\ee supposing that all subsystems are equal,
\be\label{03} \\
X(N)=NX(1) \\
\ee with the notation $X(N)\equiv{X(\sum_{i=1}^nA_{i})}$ and
$X(1)\equiv{X(A_{1})}$. Another related concept is {\it
extensivity }which corresponds to a weaker demand, namely that of,
\be\label{04}\\
\lim_{N\rightarrow\infty}\frac{|X(N)|}{N}<\infty\\
\ee Clearly, all quantities which are additive, are also
extensive, whereas the opposite is not necessarily true. In other
words, extensivity is defined as additivity when
$N\rightarrow\infty$. Of course, there are quantities that are
neither additive nor extensive. They are called {\it nonextensive
}. Boltzmann-Gibbs $(BG)$ statistical mechanics is based on the
entropy
\be\label{05} \\
S_{BG}\equiv{-k\sum_{i=1}^Wp_{i}\ln{p_{i}}}\\
\ee with
\be\label{06}\\
\sum_{i=1}^Wp_{i}=1\\
\ee where $p_{i}$ is the probability associated with the $i^{th}$
microscopic state of the system and $k$ is Boltzmann constant.
From now on, and without loss of generality, we shall take $k$
equal to unity.\\
Nonextesive statistical mechanics, first introduced by C. Tsallis
in 1988 \cite{2,3,4}, is based on the so-called 'nonextensive'
entropy $S_{q}$ defined as follows:
\be\label{07}\\
S_{q}\equiv\frac{1-\sum_{i=1}^Wp_{i}^q}{q-1}\\
\ee As we see this entropy depends on parameter $q$. Afterwards,
some other entropies were suggested depending on one
parameter \cite{5,6,7,8} .\\
Recently, an entropy was introduced \cite{9*,9} that depends on
two parameters, and in some special limits recovers other
entropies that had been introduced previously. That is
\be\label{08} S_{k,r}\equiv-\sum_{i=1}^Wp_{i}\ln_{k,r}p_{i} \ee
with
\be\label{09}\\
\ln_{k,r}(x)=x^{r}\frac{x^{k}-x^{-k}}{2k}\\
\ee The concept of extensivity has been investigated mostly for
systems with no correlation, namely independent systems. In that
case, the probabilities belong to the composition system are
defined as the product of the probabilities in each subsystem. If
the composition law is {\it not} explicitly indicated, it is
tacitly assumed that systems are statistically independent. In
that case, for two systems A and B, it immediately follows that
\be\label{010}\\
S_{BG}(A+B)=S_{BG}(A)+S_{BG}(B)\\
\ee hence, BG-entropy is additive and also extensive, but for
q-entropy we have
\be\label{011}\\
S_{q}(A+B)=S_{q}(A)+S_{q}(B)+(1-q)S_{q}(A)S_{q}(B)\ee hence,
q-entropy is nonextensive for $q\neq{1}$. In \cite{10} Tsallis
has illustrated the remarkable changes that occur when $A$ and
$B$ are specially correlated. Indeed, he has shown that in such
case
\be\label{012}\\
S_{q}(A+B)=S_{q}(A)+S_{q}(B)\\
\ee for the appropriate value of $q$ (hence extensive), whereas
\be\label{013}\\
S_{BG}(A+B)\neq{S_{BG}(A)+S_{BG}(B)}\\
\ee hence BG-entropy isn't extensive in the case of correlated
systems \cite{14}.\\
This paper is organized as follows. In sec. 2, the nonextensivity
of $S_{k,r}$ is discussed, where the extensivity of BG-entropy is
recovered in an special limit. In sec. 3, we investigate how to
interpret entropy $S_{k,r}$ extensive and finally in sec. 4
extensivity of entropy with canonic ensemble is discussed when we
have correlated subsystems.

\section{nonextensivity of $S_{k,r}$ in the case of independent systems}
As said, the entropy $S_{k,r}$ Eqs. ({\ref{08}}) and ({\ref{09}})
is more general than the other entropies introduced previously and
in some special limits recovers them. We prove that this entropy
is nonextensive in the case of independent subsystems. Supposing two
independent subsystems $A$ and $B$, for the probability in the
composite system $A+B$ we have
\begin{equation}\label{11}\\
p_{ij}^{A+B}=p_{i}^{A}p_{j}^{B}\hspace{.4cm} \forall{(i,j)}
\end{equation}
with the definitions
\be\label{13}\\
S_{k,r}(A)\equiv-\sum_{i=1}^{W_A}p_{i}^A\ln_{k,r}p_{i}^A\\
\ee
\be\label{12}\\
S_{k,r}(A+B)\equiv-\sum_{i=1}^{W_A}\sum_{j=1}^{W_{B}}p_{ij}^{A+B}\ln_{k,r}p_{ij}^{A+B}\\
\ee By adding and subtracting the phrase $p_i^{r+k+1}p_j^{r-k+1}$
 and using Eq. (\ref{09}), we can find
\begin{equation}\label{14}
S_{k,r}(A+B)=\sum_{j=1}^{W_B}p_j^{r-k+1}S_{k,r}(A)+\sum_{i=1}^{W_A}p_i^{r+k+1}S_{k,r}(B)\\
\end{equation}
As we see
\be\label{15}\\
S_{k,r}(A+B)\neq{S_{k,r}(A)+S_{k,r}(B)}\\
\ee hence $S_{k,r}$ isn't extensive in general. However, one may
choose some special range of parameters where Eq. (\ref{14}) is
extensive. We study the
extensivity of $S_{k,r}$ in some special limits.\\

\subsection{q-entropy(Tsallis entropy) }
The q-logarithm that is usually used is
\begin{equation}\label{16}\\
\ln_{q}(x)\equiv\frac{x^{1-q}-1}{1-q}\\
\end{equation}
With this logarithm the q-entropy is defined as
\begin{equation}\label{17}\\
S_q(p)\equiv-\sum_{i=1}^{W}p_{i}^q\ln_{q}p_{i}\\
\end{equation}
by choosing $r=k$ and $q=1+2k$ in Eqs. (\ref{08}) and (\ref{09}),
one has
\begin{equation}\label{18}\\
S_q(p)\equiv-\sum_{i=1}^{W}p_{i}\ln_{q}p_{i}\\
\end{equation}
where
\begin{equation}\label{19}\\
\ln_{q}(x)\equiv\frac{x^{q-1}-1}{q-1}\\
\end{equation}
which gives an equivalent entropy to Eq. (\ref{17}). For
$\exp_q(x)$ it is obtained
\begin{equation}\label{110}\\
{\exp}_{q}(x)\equiv{[\hspace{.08cm}1+(q-1)x\hspace{.08cm}]^{\frac{1}{q-1}}}\\
\end{equation}
In the limit $r=k$ and $q=1+2k$ from Eq. (\ref{14}), one recovers
\begin{equation}\label{111}
S_{q}(A+B)=\sum_{j=1}^{W_B}p_jS_{q}(A)+\sum_{i=1}^{W_A}p_i^{q}S_{q}(B)\\
=S_{q}(A)+S_{q}(B)+(1-q)S_{q}(A)S_{q}(B)\\
\end{equation}
that is the familiar expression for nonextensivity of q-entropy.
In the limit $q\rightarrow{1}$ extensivity of BG-entropy is
obtained.

\subsection{k-entropy}
The k-entropy introduced in \cite{7,8} is
\begin{equation}\label{112}\\
S_k(p)\equiv-\sum_{i=1}^{W}p_{i}\ln_{k}p_{i}\\
\end{equation}
where
\begin{equation}\label{113}\\
\ln_{k}(x)\equiv\frac{x^k-x^{-k}}{2k}\\
\end{equation}
It is clear that we can recover k-logarithm from Eq. (\ref{09}) in
the limit $r\rightarrow{0}$. For $\exp_k(x)$ we have
\begin{equation}\label{114}\\
\exp_k(x)\equiv{(\sqrt{1+k^2x^2}+kx)^{\frac{1}{k}}}\\
\end{equation}
In that limit Eq. (\ref{14}) results in \ba
&&S_{k}(A+B)=\sum_{j=1}^{W_B}p_j^{-k+1}S_{k}(A)+\sum_{i=1}^{W_A}p_i^{k+1}S_{k}(B)\nn\\
&&\hspace{2cm}\neq{S_{k}(A)+S_{k}(B)}\label{115} \ea that ensures
the nonextnsivity of the k-entropy. It is clear that in the limit
$k\rightarrow{0}$ the extensivity  of BG-entropy is obtained
again.
\section{How to interpret the entropy $S_{k,r}$ extensive}
Suppose, we have $N$ subsystems $(A_{1},A_{2},...,A_{N})$. We
define the probabilities in the composite system
$p_{i_1i_2...i_N}^{A_1+A_2+...+A_N}$ that satisfy the condition
\begin{equation}\label{21}
\sum_{i_1i_2...i_N}p_{i_1i_2...i_N}^{A_1+A_2+...+A_N}=1
\end{equation}
and marginal probabilities as follows
\begin{equation}\label{22}
p_{i_s}^{A_s}\equiv\sum_{i_1i_2...i_{s-1}i_{s+1}i_N}p_{i_1i_2...i_N}^{A_1+A_2+...+A_N}
\end{equation}
If $p_{i_1i_2...i_N}^{A_1+A_2+...+A_N}$ also satisfies the
condition
\begin{equation}\label{23}
p_{i_1i_2...i_N}^{A_1+A_2+...+A_N}=\exp_{k,r}(\hspace{.07cm}\sum_{s=1}^{N}\ln_{k,r}p_{i_s}^{A_s})
\end{equation}
then for the entropy of the composite system with the definition
\begin{equation}\label{24}
S_{k,r}(\sum_{s=1}^{N}A_s)\equiv-\sum_{i_1i_2...i_N}p_{i_1i_2...i_N}^{A_1+A_2+...+A_N}\\
\ln_{k,r}p_{i_1i_2...i_N}^{A_1+A_2+...+A_N}
\end{equation}
we have \ba
&&S_{k,r}(\sum_{s=1}^{N}A_s)=-\sum_{i_1i_2...i_N}p_{i_1i_2...i_N}^{A_1+A_2+\ldots+A_N}
\ln_{k,r}[\exp_{k,r}(\hspace{.07cm}\sum_{s=1}^{N}\ln_{k,r}p_{i_s}^{A_s})]\nn\\
&&\hspace{2.1cm}=-\sum_{i_1i_2...i_N}p_{i_1i_2...i_N}^{A_1+A_2+\ldots+A_N}\sum_{s=1}^{N}\ln_{k,r}p_{i_s}^{A_s}\nn\\
&&\hspace{2.1cm}=-\sum_{s=1}^N\sum_{i_s}p_{i_s}^{A_s}\ln_{k,r}p_{i_s}^{A_s}
=\sum_{s=1}^NS_{k,r}(A_s)\label{25} \ea It is useful at this
point to connect the present problem to some generalized algebra
which have been discussed by many authors. We use the
product introduced in \cite{9}. It is defined as follows:\\
\begin{equation}\label{26}
x\otimes_{k,r}{y}\equiv\exp_{k,r}\big(\ln_{k,r}(x)+\ln_{k,r}(y)\big)
\end{equation}
hence we can write (\ref{23}) as
\ba
&&p_{i_1i_2...i_N}^{A_1+A_2+...+A_N}=\exp_{k,r}(\hspace{.07cm}\sum_{s=1}^{N}\ln_{k,r}p_{i_s}^{A_s})\nn\\
&&\hspace{2.5cm}=p_{i_1}^{A_1}\otimes_{k,r}{p}_{i_2}^{A_2}{\otimes}_{k,r}^{}\ldots{\otimes}_{k,r}p_{i_N}^{A_N}
\label{27} \ea So extensivity of the entropy is satisfied if we
use logarithm, exponential and also the product based on
(\ref{26}). In the limit $k\rightarrow{0}$ and $r\rightarrow{0}$
(BG-limit), the usual product is recovered and (\ref{27})
describes the probability of composite system in the case of
independent subsystems and  also extensivity of BG-entropy in
that case which is expected. Eq. (\ref{23}) is a very special
correlation for subsystems which leads to extensivity of entropy.
however, it is possible to define a general correlation among
subsystems so that the entropy remains extensive. Consider the
following relation
\begin{equation}\label{29}
{\widetilde{p}}_{i_1i_2...i_N}^{A_1+A_2+...+A_N}{\equiv}\hspace{.15cm}{{{p_{i_1}}}^{A_1}}{\otimes}_{k,r}
p_{i_2}^{A_2}{\otimes}_{k,r}\ldots{\otimes}_{k,r}p_{i_N}^{A_N}
\end{equation}
where $p_{i_s}^{A_s}$s are the probabilities of each subsystem,
but ${\widetilde{p}}_{i_1i_2...i_N}^{A_1+A_2+...+A_N}$s are not
necessarily represent the joint probabilities. Now the sum of
subsystem entropies can be written as \ba
&&\sum_{s=1}^NS_{k,r}(A_s)
=-\sum_{s=1}^N\sum_{i_s}p_{i_s}^{A_s}\ln_{k,r}p_{i_s}^{A_s}
=-\sum_{i_1i_2...i_N}p_{i_1i_2...i_N}^{A_1+A_2+\ldots+A_N}\sum_{s=1}^{N}\ln_{k,r}p_{i_s}^{A_s}\nn\\
&&\hspace{2.1cm}=-\sum_{i_1i_2...i_N}p_{i_1i_2...i_N}^{A_1+A_2+\ldots+A_N}
\ln_{k,r}[\exp_{k,r}(\hspace{.07cm}\sum_{s=1}^{N}\ln_{k,r}p_{i_s}^{A_s})]\nn\\
&&\hspace{2.1cm}=-\sum_{i_1i_2...i_N}p_{i_1i_2...i_N}^{A_1+A_2+\ldots+A_N}
\ln_{k,r}{\widetilde{p}}_{i_1i_2...i_N}^{A_1+A_2+...+A_N}
\label{210} \ea So entropy is extensive if
\begin{eqnarray}
&&S_{k,r}(\sum_{s=1}^{N}A_s)=-\sum_{i_1i_2...i_N}p_{i_1i_2...i_N}^{A_1+A_2+\ldots+A_N}
\ln_{k,r}{p}_{i_1i_2...i_N}^{A_1+A_2+...+A_N}\nn\\
&&\hspace{2.1cm}=-\sum_{i_1i_2...i_N}p_{i_1i_2...i_N}^{A_1+A_2+\ldots+A_N}
\ln_{k,r}{\widetilde{p}}_{i_1i_2...i_N}^{A_1+A_2+...+A_N}\label{28}
\end{eqnarray}
It is clear that
${\widetilde{p}}_{i_1i_2...i_N}^{A_1+A_2+...+A_N}$ and
${p}_{i_1i_2...i_N}^{A_1+A_2+...+A_N}$ can be related to each
other by the following relations
\begin{equation}\label{211}
\ln_{k,r}{p}_{i_1i_2...i_N}^{A_1+A_2+...+A_N}-\ln_{k,r}{\widetilde{p}}_{i_1i_2...i_N}^{A_1+A_2+...+A_N}
=\phi_{i_1i_2...i_N}
\end{equation}
\begin{equation}\label{213}
\hspace{.9cm}\sum_{i_1i_2...i_N}p_{i_1i_2...i_N}^{A_1+A_2+...+A_N}\phi_{i_1i_2...i_N}=0
\end{equation}
where $\phi_{i_1i_2...i_N}$ are arbitrary functions with
(\ref{213}) as a constraint. Eqs. (\ref{29}) and (\ref{211})
result in
\begin{equation}\label{214}
p_{i_1i_2...i_N}^{A_1+A_2+...+A_N}=\exp_{k,r}(\sum_{s=1}^N\ln_{k,r}p_{i_s}^{A_s}+\phi_{i_1i_2...i_N})
\end{equation}
In the Tsallis limit Eq. (\ref{214}) can be written as (by using
(\ref{19}) and (\ref{110}))
\begin{equation}\label{215}
p_{i_1i_2...i_N}^{A_1+A_2+...+A_N}={\Big[1-N+(q-1)\phi_{i_1i_2...i_N}
+\sum_{s=1}^{N}{(p_{i_s}^{A_s})}^{q-1}\Big]}^{\frac{1}{q-1}}
\end{equation}
which is equivalent to the Tsallis proposal for the joint
probabilities \cite{10} if we choose
\begin{equation}\label{216}
\phi_{i_1i_2...i_N}^{(q)}=(q-1)\phi_{i_1i_2...i_N}
\end{equation}
Consider two subsystems A and B where the probabilities of
composite system and each subsystem are shown in the following
table

\begin{center}
\begin{tabular}{c||c|c|c}

$A\setminus{B}$ & 1 & 2 &  \\ \cline{1-4}\hline\hline\
  1 & $p_{11}^{A+B}$ & $p_{12}^{A+B}$ & $p_1^A$ \\ \cline{1-4}\hline
  2 &  $p_{21}^{A+B}$ & $p_{22}^{A+B}$ & $1-p_1^A$ \\ \cline{1-4} \hline
   & $p_1^B$ & $1-p_1^B$ & 1 \\
\end{tabular}
\end{center}
with the following relations \ba
&&p_{11}^{A+B}+p_{12}^{A+B}=p_1^A\label{218}\\
&&p_{21}^{A+B}+p_{22}^{A+B}=p_2^A=1-p_1^A\label{219}\\
&&p_{11}^{A+B}+p_{21}^{A+B}=p_1^B\label{220}\\
&&p_{12}^{A+B}+p_{22}^{A+B}=p_2^B=1-p_1^B\label{221}\ea and also
a constraint (\ref{213})
\begin{equation}\label{222}
p_{11}^{A+B}\phi_{11}+p_{12}^{A+B}\phi_{12}+p_{21}^{A+B}\phi_{21}+p_{22}^{A+B}\phi_{22}=0
\end{equation}
Using Eq. (\ref{214}), it is possible to write Eqs. (\ref{218}) to
(\ref{221}) in terms of $p_1^A, p_2^A, \phi_{11}, \phi_{12},
\phi_{21}$ and $\phi_{22}$. So $\phi_{ij}$s can be determined.
For simplicity we use Tsallis limit and so (\ref{215}) for the
probabilities of  the composite system. We also assume that both
subsystems $A$ and $B$ are equal, namely $p_1^A=p_1^B=p$. So we
have \ba &&p_{11}^{A+B}={[2p^{q-1}+(q-1)\phi_{11}
-1]}^{\frac{1}{q-1}}\label{223}\\
&&p_{12}^{A+B}={[p^{q-1}+{(1-p)}^{q-1}+(q-1)\phi_{12}
-1]}^{\frac{1}{q-1}}\label{224}\\
&&p_{21}^{A+B}={[{(1-p)}^{q-1}+p^{q-1}+(q-1)\phi_{21}
-1]}^{\frac{1}{q-1}}\label{225}\\
&&p_{22}^{A+B}={[2{(1-p)}^{q-1}+(q-1)\phi_{22}
-1]}^{\frac{1}{q-1}}\label{226}\ea By substituting Eqs.
(\ref{223}) to (\ref{226}) in (\ref{218}) to (\ref{222}), we
obtain $\phi_{12}=\phi_{21}$ and so \ba &&
{[2p^{q-1}+(q-1)\phi_{11} -1]}^{\frac{1}{q-1}}
+{[p^{q-1}+{(1-p)}^{q-1}+(q-1)\phi_{12}
-1]}^{\frac{1}{q-1}}\nn\\&&\hspace{11.5cm}=p\label{227}\\
&&{[p^{q-1}+{(1-p)}^{q-1}+(q-1)\phi_{12} -1]}^{\frac{1}{q-1}}
+{[2{(1-p)}^{q-1}+(q-1)\phi_{22}
-1]}^{\frac{1}{q-1}}\nn\\&&\hspace{11cm}=1-p\label{228}\\
&&\phi_{11}{[2p^{q-1}+(q-1)\phi_{11}
-1]}^{\frac{1}{q-1}}+2\phi_{12}{[p^{q-1}+{(1-p)}^{q-1}+(q-1)\phi_{12}
-1]}^{\frac{1}{q-1}}\nn\\
&&+\phi_{22}{[2{(1-p)}^{q-1}+(q-1)\phi_{22}
-1]}^{\frac{1}{q-1}}=0\label{229}\ea With a given value of $q$,
 above equations can be solved and
one may obtain $\phi_{11}(p),\phi_{12}(p)=\phi_{21}(p)$ and
$\phi_{22}(p)$. A few typical $(q,\phi_{11}(1/2),\phi_{12}(1/2))$
points are:
$\\(0.4,-7.57873,0.71845), (0.5,-4.20199,0.62528), (0.6,-2.5339,0.53573), \\
(0.7,-1.56656,0.44851), (0.8,-0.94037,0.35940),
(0.9,-0.49032,0.25676)$ where it should be noted that for $p=1/2$
symmetries of equations ensures that $\phi_{11}(p)=\phi_{22}(p)$.
We will investigate more numerical estimates for two-parameter
entropies in another paper.
\section{Extensive entropy in the case of correlated subsystems, a constraint approach}
In this section, a new approach is used where the condition
(\ref{213}) is entered to the entropy as a constraint and then the
entropy is maximized . Parallel to what is done in \cite{9}, we
introduce the entropy in the composite system as
\begin{equation}\label{34}
S(\sum_{s=1}^{N}A_s)\equiv-\sum_{i_1i_2...i_N}p_{i_1i_2...i_N}^{A_1+A_2+...+A_N}\\
\Lambda{(p_{i_1i_2...i_N}^{A_1+A_2+...+A_N})}
\end{equation}
where $\Lambda{(x)}$ is a generalization of the logarithm. We have
the constraints \ba
&&\sum_{i_1i_2...i_N}p_{i_1i_2...i_N}^{A_1+A_2+...+A_N}=1\label{21*}\\
&&\sum_{i_1i_2...i_N}p_{i_1i_2...i_N}^{A_1+A_2+...+A_N}E_{i_1i_2...i_N}^{A_1+A_2+...+A_N}=U\label{35}\\
&&\sum_{i_1i_2...i_N}p_{i_1i_2...i_N}^{A_1+A_2+...+A_N}\phi_{i_1i_2...i_N}=0\label{36}
\ea For simplicity we use the notation $\{i\}$ instead of
$\{i_1i_2...i_N\}$. Then the entropic functional can be
introduced as
\be\label{37}\\
{\cal F}[p]=S(p)-{\beta}^{\prime}\\
\Big(\sum_{\{i\}}p_{\{i\}}-1\Big)-\beta\Big(\sum_{\{i\}}p_{\{i\}}E_{\{i\}}-U\Big)
-{\beta}''\Big(\sum_{\{i\}}p_{\{i\}}\phi_{\{i\}}\Big)\\
\ee where $\beta$, ${\beta}'$ and ${\beta}''$ are Lagrange
multipliers and it has been supposed that $\phi_{\{i\}}$ isn't an
explicit function of $p_{\{i\}}$. If ${\cal F}[p]$ in Eq.
(\ref{37}) is stationary for variations of the probabilities
$p_{\{j\}}$,
\begin{equation}\label{38}
\frac{\delta}{{\delta}p_{\{j\}}}{\cal F}[p]=0
\end{equation}
one finds
\begin{equation}\label{39}
\frac{d}{dp_{\{j\}}}\big[\hspace{.05cm}p_{\{j\}}\Lambda(p_{\{j\}})\big]=-\beta(E_{\{j\}}-\mu-{\mu}'\phi_{\{j\}})
\end{equation}
where $\mu=-{\beta}'/\beta$ and ${\mu}'=-{\beta}''/\beta$.\\
Without loss of generality, we can express the probability
distribution $p_j$ as
\begin{equation}\label{310}
p_{\{j\}}=\alpha{\bf
\varepsilon}\Big(-\frac{\beta}{\lambda}(E_{\{j\}}-\mu-{\mu}'\phi_{\{j\}})\hspace{.05cm}\Big)
\end{equation}
where $\alpha$ and $\lambda$ are two arbitrary, real and positive
constants, and $\varepsilon(x)$ an invertible function that can be
a generalization of, and in some limit reduce to, the exponential
function. If we require that $\varepsilon(x)$ be the inverse of
$\Lambda(x)$, Eqs. (\ref{39}) and (\ref{310}) result in
\begin{equation}\label{311}
\frac{d}{dp_{\{j\}}}\big[\hspace{.05cm}p_{\{j\}}\Lambda(p_{\{j\}})\big]=\lambda{{\bf
\varepsilon}}^{-1}\big(\frac{p_{\{j\}}}{\alpha}\big)
\end{equation}
that can be rewritten as \cite{9}
\begin{equation}\label{312}
\frac{d}{dx}\big[\hspace{.05cm}x\Lambda(x)\big]=\lambda{{\bf
\varepsilon}}^{-1}\big(\frac{x}{\alpha}\big)
\end{equation}
So, for $\Lambda{(x)}$ we have
\begin{equation}\label{313}
\Lambda(x)=\ln_{k,r}(x)=x^r\frac{x^k-x^{-k}}{2k}
\end{equation}
and the constants $\alpha$ and $\lambda$ can be expressed in
terms of $k$ and $r$ \ba
&&\alpha={\big(\frac{1+r-k}{1+r+k}\big)}^{1/(2k)}\label{313&}\\
&&\lambda=\frac{{(1+r-k)}^{{(r+k)}/{(2k)}}}{{(1+r+k)}^{{(r-k)}/{(2k)}}}\label{313&&}
\ea Eq. (\ref{313}) indicates that by imposing the condition
(\ref{36}), the definition of logarithm dose not change and the
only thing we must change is the definition of probability in the
composite system.\\
It is useful here to interpret each subsystem separately. By
imposing the conditions \ba
&&\sum_{i_s}p_{i_s}^{A_s}=1\label{314}\\
&&\sum_{i_s}p_{i_s}^{A_s}E_{i_s}^{A_s}=U_s\label{315} \ea For the
subsystems, the entropic functional will be \be\label{317}
{\cal F}_s[p]=S_s(p)-{{\beta}^{\prime}}_s\\
\Big(\sum_{i_s}p_{i_s}-1\Big)-{\beta}_s\Big(\sum_{i_s}p_{i_s}E_{i_s}-U_s\Big)
\ee and by maximizing the entropic functional in the way similar
to the case of composite system, we obtain
\begin{equation}\label{318}
p_{i_s}=\alpha\exp_{k,r}
\Big(-\frac{{\beta}_s}{\lambda}(E_{i_s}-\mu_s)\hspace{.05cm}\Big)
\end{equation}
where $\mu_s=-{{\beta}'}_s/{\beta}_s$, $\exp_{k,r}(x)$ is inverse
function of $\ln_{k,r}(x)$ and $\alpha$
and $\lambda$ are defined in Eqs. (\ref{313&}) and (\ref{313&&}). \\
Using (\ref{310}) and (\ref{318}), Eq. (31) can be written as
\begin{equation}\label{319}
\ln_{k,r}[\alpha\exp_{k,r}(-\frac{\beta}{\lambda}(E_{\{j\}}-\mu-{\mu}'\phi_{\{j\}})\hspace{.05cm}\big)]=
\sum_{s=1}^N\ln_{k,r}[\alpha\exp_{k,r}(-\frac{\beta_s}{\lambda}(E_{i_s}-\mu_s)\hspace{.05cm}\big)]
\end{equation}
Where parameters $\phi_{\{j\}}$ can be used to ensure extensivity
of  the two-parameter entropies. From Eq. (\ref{319}), it is
clear that extensivity of entropy dose not necessarily ensures
extensivity of energy ( For a discussion in the case of q-entropy
see \cite{13} ) . In the Boltzmann-Gibbs limit Eq. (\ref{319})
becomes

\begin{equation}\label{319*}
{\beta}(E_{\{j\}}-\mu-{\mu}'\phi_{\{j\}})=\sum_{s=1}^N{\beta_s}(E_{i_s}-\mu_s)
\end{equation}
where only in a special case leads to the extensivity of energy.

\section{Probabilities and effective number of states}
 Our motivation for studying such kind of correlations and extensivity of the two-parameter entropies of
 correlated subsystems
 was the following argument by Tsallis \cite{10,11} which defines
 effective number of states. Suppose that the probability
 distribution in phase space is uniform within a volume W and also
 $S_q$ is given by
\begin{equation}\label{*1}
 S_q=\ln_qW
\end{equation}
With the help of q-product \cite{12} defined as \ba
&&x\otimes_qy\equiv\exp_q(\ln_qx+\ln_qy)\nn\\
&&\hspace{1.2cm}=(x^{1-q}+y^{1-q}-1)^{1/{1-q}}\label{*2} \ea
 it is possible to interpret $S_q$ extensive. Supposing
that $W_A$ and $W_B$ be the number of states for subsystems A and
B. Equation
\begin{equation}\label{*3}
 W_{A+B}^{eff}\equiv{W_A}\otimes_qW_B
\end{equation}
can be interpreted as a definition for effective number of states
for the system $A+B$. Definition (\ref{*2}) ensures that
\begin{equation}\label{*4}
\ln_qW_{A+B}^{eff}=\ln_q{W_A}\otimes_qW_B=\ln_qW_A+\ln_qW_B
\end{equation}
Eq. (\ref{*4}) shows extensivity of the entropy (\ref{*1}). If we
suppose
\begin{equation}\label{*5}
W^{A_1}=W^{A_2}=...=W^{A_N}=1/p
\end{equation}
the probability in the composite system will be
\begin{equation}\label{*6}
(1/p_{i_1i_2...i_N}^{A_1+A_2+\ldots+A_N})=(1/p)\otimes_q(1/p)\otimes_q...\otimes_q(1/p)
\end{equation}
and hence
\begin{equation}\label{*7}
p_{i_1i_2...i_N}^{A_1+A_2+\ldots+A_N}=p\otimes_{2-q}p\otimes_{2-q}...\otimes_{2-q}p
\end{equation}
where (\ref{*7}) is obtained from (\ref{*6}) by the properties of
q-product. At this point it is appropriate to use the following
q-product which is used in this paper \ba
&&x\otimes_{q'}y\equiv\exp_{q'}(\ln_{q'}x+\ln_{q'}y)\nn\\
&&\hspace{1.2cm}=(x^{q'-1}+y^{q'-1}-1)^{1/{q'-1}}\label{*8} \ea
comparing Eq. (\ref{*8}) with Eq. (\ref{*2}) shows that $q'=2-q$.
By our q-product Eq. (\ref{*7}) can be written as
\begin{equation}\label{*9}
p_{i_1i_2...i_N}^{A_1+A_2+\ldots+A_N}=p\otimes_{q}p\otimes_{q}...\otimes_{q}p
\end{equation}
This is a hinting point to define the probability of composite
system in terms of the probabilities of subsystems by a
generalized $(k,r)$-product.

\section{conclusion}
In this paper, it is shown that two-parameter entropies $S_{k,r}$
are not in general extensive. A formulation is given where by
$(k,r)$-products of subsystem probabilities one may obtain joint
probabilities involving some functions $\phi_{i_1i_2...i_N}$.
Demanding extensivity of the entropy imposes some constraints on
 $\phi_{i_1i_2...i_N}$s and so joint probabilities are
identified. We believe this is the most general representation
for obtaining extensive entropies in the case of correlated
subsystems.

\end{document}